\documentclass[12pt]{iopart}
\usepackage{iopams}  
\usepackage{braket}
\usepackage{graphicx}
\usepackage{relsize}

\newcommand{\be}{\begin{equation}}
\newcommand{\ee}{\end{equation}}
\newcommand{\gs}{\ensuremath{\sigma}}
\newcommand{\gd}{\ensuremath{\delta}}

\begin{document}

\title[Non-Markovian Microscopic Depolarizing Channel]{Simple Non-Markovian Microscopic Models for the Depolarizing Channel of a Single Qubit}
% Simple Microscopic Models of One-Qubit Noisy Channels are Non-Markovian 

\author{K. M. Fonseca Romero$^1$ and R. Lo Franco$^2$}
\address{$^1$Universidad Nacional de Colombia - Bogot\'a, Facultad de Ciencias,  Departamento de F{\'isica}, Grupo de {\'O}ptica e Informaci{\'o}n Cu{\'a}ntica, Carrera 30 Calle 45-03, C.P. 111321, Bogot{\'a}, Colombia\\
$^2$Dipartimento di Fisica, Universit\`a di Palermo, 
via Archirafi 36, 90123 Palermo, Italy}
\ead{kmfonsecar@unal.edu.co}

\begin{abstract}
The archetypal one-qubit noisy channels ---depolarizing, phase-damping and amplitude-damping channels--- describe both Markovian and non-Markovian evolution. Simple microscopic models for the depolarizing channel, both classical and quantum, are considered. Microscopic models which describe phase damping and amplitude damping channels are briefly reviewed. 
\end{abstract}

 % Decoherence; open systems; quantum statistical methods.
\submitto{\PS}
%\maketitle

\section{Introduction}
The decoherence process followed by a given quantum physical system is a consequence of its unavoidable interaction with the environment. The ensuing non-unitary dynamics is usually modelled by a (generalized) master equation for the system's density operator. In the literature of open quantum systems, master equations are usually called Markovian if they did not involve an explicit time integration. Otherwise, they are termed non-Markovian. These designations, however, do not agree with the mathematical definitions of Markovian and non-Markovian processes, as was shown long ago \cite{Oppenheim1965PhysRev138.B1007}.
Moreover, non-Markovian master equations can be rewritten in time-local form (without explicit time integration), as it was recently shown \cite{Andersson2007, Chruscinski2010}. Thus, an alternative, more specific criterion, will be used in this paper to characterize non-Markovian dynamics. The dynamics of a system is non-Markovian if it can be described by a master equation of generalized Lindblad form, with temporarily negative decay rates \cite{Breuer2004, Andersson2010, Laine2012}.

In order to estimate the effects of nonunitary dynamics on different quantum information processing protocols, it is customary to use several archetypal quantum operations on a qubit \cite{Nielsen2000Q}: the depolarizing, phase-damping (or phase-flip) and amplitude-damping channels. They are usually assumed  to be Markovian, with constant decay rates, in theoretical and experimental analyses \cite{Almeida2007Science5824.579}. However, the description of decoherence in solid state systems, the most promising scalable realizations of quantum processors, often needs to be non-Markovian \cite{John1995PRL74.3419,Quang1997PRL79.5238,deVega2005PRA71.023812,Florescu2004PRA69.013816,Bellomo2008PRA78.060302,
Bellomo2010PST140.014014,Bellomo2010PRA81.062309,Bellomo2011PST143.014004}. Thus, it is important to determine the generality of the analyses using the above mentioned noisy channels, and if the usual Markovian assumption is unnecessary. The purpose of this manuscript is to show, explicitly, that the archetypal one-qubit noisy channels, described before, generally describe non-Markovian dynamics.

This manuscript is organized as follows. In section \ref{sec:MLK}, a brief review is made of Markov and Lindblad master equations, Kraus operator sum representation and the usual one qubit noisy channels. We consider quantum models of noise for one-qubit depolarizing (section~\ref{sec:microdepolarizing}), dephasing (section~\ref{sec:microdephasing}) and amplitude damping (section~\ref{sec:microdamping}) channels, and classical models of noise for one-qubit depolarizing (section~\ref{sec:macrodepolarizing}) and dephasing (section~\ref{sec:macrodephasing}) channels. These models are simple yet (generally) non-Markovian models, where decoherence of the qubit is due to the fluctuation of macroscopic variables which enter in the Hamiltonian or by the establishment of correlation betwen the system and its environment. Some conclusions are drawn in the last section of the paper. 

\section{Markov, Lindblad and Kraus} \label{sec:MLK}
A typical Markovian master equation is a dynamical equation for $\hat{\rho}_{\textrm{\relsize{-2}{S}}}(t)$, the density operator of the system of interest, which can be cast as 
\begin{equation}
 \frac{d\hat{\rho}_{\textrm{\relsize{-2}{S}}}(t)}{dt} = -\frac{i}{\hbar}\left[\hat{H},\,\hat{\rho}_{\textrm{\relsize{-2}{S}}}(t)\right]
+\mathcal{L}_{\textrm{\relsize{-2}{D}}}\hat{\rho}_{\textrm{\relsize{-2}{S}}}(t),
\end{equation} 
where $\mathcal{L}_{\textrm{\relsize{-2}{D}}}$ is a time-independent (super)operator. Not all Markovian master equations, however, can be written in this way, as we will see below. Markovian master equations are important tools to model open quantum systems, not only due to their mathematical simplicity, but also because they capture the physical behavior of many important systems, such as open QED (Quantum Electrodynamics) systems \cite{CohenTannoudji1992A,Gardiner1985H,Scully1997Q,Walls1994Q}. Many Markovian master equations can be derived from system-environment models employing the Born-Markov approximation \cite{Louisell1973Q}, which relies on a series of assumptions which are not always satisfied \cite{Breuer2002}: weak system-environment coupling, separable total density operator, bath correlation time much smaller than the relaxation time of the system, and unperturbed transition times of the system much smaller than its relaxation time. The next to last assumption implies that the environment is 
infinite. Thus, loosely 
speaking, Markovian behavior is typical of systems which interact weakly with infinite environments. Sometimes, Markovian master equations present unphysical behavior, because they lead to negative density operators \cite{Suarez1992JChemPhys7.5101,Gnutzmann1996ZfPB101.263}.

Master equation of the Lindblad form \cite{Lindblad1976CommMathPhys48.119,Gorini1976JMP17.821}, which preserve the fundamental properties of density operators (non-negativity, unit trace and hermicity), are also Markovian. The standard form of Lindblad master equations \cite{Lindblad1976CommMathPhys48.119} for the density operator of the system of interest, $\hat{\rho}_{\textrm{\relsize{-2}{S}}}(t)$, is
\begin{eqnarray}\label{eq:mastereq}\frac{d\hat{\rho}_{\textrm{\relsize{-2}{S}}}(t)}{dt}=&-\frac{i}{\hbar}\left[\hat{H},\,\hat{\rho}_{\textrm{\relsize{-2}{S}}}(t)\right]\\ & \nonumber
+\sum_i\left(2\,\hat{L}_i\,\hat{\rho}_{\textrm{\relsize{-2}{S}}}(t)\,\hat{L}_i^\dagger-\hat{L}_i^\dagger\,\hat{L}_i\,\hat{\rho}_{\textrm{\relsize{-2}{S}}}(t)-\hat{\rho}_{\textrm{\relsize{-2}{S}}}(t)\,\hat{L}_i^\dagger\,\hat{L}_i\right),\end{eqnarray}
where $\hat{H}$ is a Hamiltonian and $\{\hat{L}_j\}$ are (possibly non-hermitian) operators, usually known as Lindblad operators. For a given physical process the decomposition (\ref{eq:mastereq}) is not unique. However, it is convenient to interpret the first term of the right hand side of (\ref{eq:mastereq}) as the unitary evolution in the absence of interaction with the environment, and the second term as the environment-system coupling.
Despite their mathematical properties, under special conditions Lindblad master equations can also exhibit unphysical behavior, like the unbounded increase of the system's energy \cite{Berredo1998PhysScripta57.533}.

One of the alternatives to the use of generalized master equations to describe open quantum systems is the Feynman-Vernon influence functional \cite{Feynman1963AnnPhys24.118}. Other, which was derived by Kraus \cite{Kraus1983S}  employing the idea of complete positivity, is known as the ``operator-sum representation''  of the quantum dynamics of an open  system. The Kraus representation of a decoherence process 
\begin{equation}\label{Kraus}
\hat{\rho}_{\textrm{\relsize{-2}{S}}}(t)=\sum_i\hat{E}_i(t)\,\hat{\rho}_{\textrm{\relsize{-2}{S}}}(0)\,\hat{E}_i^\dagger(t),
\end{equation}
where the Kraus operators $\hat{E}_i(t)$ satisfy the condition of probability conservation $\sum_i\hat{E}_i^\dagger(t)\,\hat{E}_i(t)=\hat{{I}}$, gives the evolution of the density matrix of the system of interest. 

The operator sum representation and Lindblad-form master equations can be related. In effect, if a closed quantum system, initially in a product state $\hat{\rho}_{\textrm{\relsize{-2}{T}}}(0)=\hat{\rho}_{\textrm{\relsize{-2}{S}}}(0)\otimes \hat{\rho}_{\textrm{\relsize{-2}{E}}}(0)$, comprises two interacting subsystems, the system of interest $S$ and its environment $E$, the exact dynamics of the state of any of the subsystems can be put in Kraus form. If the inequality $\tau_{\textrm{\relsize{-2}{C}}}\ll \tau\ll \tau_{\textrm{\relsize{-2}{H}}}$ is satisfied, where $1/\tau_{\textrm{\relsize{-2}{C}}}$ is the cutoff frequency of the bath density of states, $\tau$ an adequate coarse-graining time scale and $\tau_{\textrm{\relsize{-2}{H}}}$ the characteristic time-scale of the hamiltonian evolution of the system, then it is possible to derive a completely-positive master equation starting from the Kraus operator-sum representation \cite{Bacon1999PhysRevA60.1944,Lidar2001ChemPhys268.35}, linking both descriptions.

The most popular quantum channels are the depolarizing, phase-damping (or phase-flip) and amplitude-damping one-qubit channels \cite{Nielsen2000Q}. We consider a qubit with orthonormal states $\ket{0}$ and $\ket{1}$. In this basis the operator $\hat\sigma_3=\ket{0}\bra{0}-\ket{1}\bra{1}$ is diagonal, and the lowering operator is given by $\hat{\sigma}_-=\ket{1}\bra{0}=\hat{\sigma}_1-i\hat{\sigma}_2=\hat{\sigma}_+^\dag$. The  depolarizing channel for  a single qubit
\begin{equation}\label{eq:depolarizing} \hat{\rho}_{\textrm{\relsize{-2}{S}}}(p)=p\,\frac{\hat{I}}{2}+(1-p)\,\hat{\rho}_{\textrm{\relsize{-2}{S}}}(0), \end{equation} describes a process where the qubit, initially in its state $\hat{\rho}_{\textrm{\relsize{-2}{S}}}(0)$, remains in this state with probability $1-p$, and changes to the maximally mixed state $\hat{I}/2$ with probability $p$. 
Another popular description of a depolarizing channel $\hat{\rho}_{\textrm{\relsize{-2}{S}}}(\tilde{p})=\frac{\tilde{p}}{3}\,\sum_i \hat{\sigma}_i \hat{\rho}_{\textrm{\relsize{-2}{S}}}(0)\hat{\sigma}_i  +(1- \tilde{p})\,\hat{\rho}_{\textrm{\relsize{-2}{S}}}(0)$ \cite{Preskill1998}, emphasizes the probability of remaining in the same state, $1-\tilde{p}$, in contrast to the probability of suffering an error. It is important to notice that $\tilde{p}= 3 p/4$, that is, these probabilities are different: while $0\leq p\leq 1$, $0\leq \tilde{p}\leq 3/4$. 
If a two-qubit mixed state is used, instead of a Bell state, in the standard teleportation protocol, a generalized depolarizing channel is obtained \cite{Bowen2001PRL87.267901}. An amplitude damping channel is obtained when a qubit interacts with a large reservoir at zero temperature, or a two-level atom in the electromagnetic vacuum emits spontaneously \cite{CohenTannoudji1992A,Gardiner1985H,Scully1997Q,Walls1994Q,Louisell1973Q}. An exciton confined to a quantum dot and coupled to phonons \cite{Roszak2006PRA73.022313}, like a spin in a magnetic field whose strength fluctuates in time, suffers a dephasing process. 

The Markovian master equations, with constant decay rates, corresponding to the depolarizing, amplitude-damping and phase-damping channels are 
\begin{eqnarray}\label{eq:depolarizingM} 
\frac{d\hat{\rho}_{\textrm{\relsize{-2}{S}}}(t)}{dt}=\frac{\gamma}{8}\sum_i\left(2\,\hat{\gs}_i\,\hat{\rho}_{\textrm{\relsize{-2}{S}}}(t)\,\hat{\gs}_i-\hat{\gs}_i^2 \hat{\rho}_{\textrm{\relsize{-2}{S}}}(t)-\hat{\rho}_{\textrm{\relsize{-2}{S}}}(t)\,\hat{\gs}_i^2\right),\\
\label{eq:amplitudedampM}\frac{d\hat{\rho}_{\textrm{\relsize{-2}{S}}}(t)}{dt} =\frac{\gamma}{2}\left(2\,\hat{\sigma}_-\,\hat{\rho}_{\textrm{\relsize{-2}{S}}}(t)\,\hat{\sigma}_+-\hat{\sigma}_+\,\hat{\sigma}_-\,\hat{\rho}_{\textrm{\relsize{-2}{S}}}(t)-\hat{\rho}_{\textrm{\relsize{-2}{S}}}(t)\,\hat{\sigma}_+\,\hat{\sigma}_-\right),\\
\frac{d\hat{\rho}_{\textrm{\relsize{-2}{S}}}(t)}{dt}=\frac{\gamma}{4}\left(2\hat{\sigma}_3\,\hat{\rho}_{\textrm{\relsize{-2}{S}}}(t) \hat{\sigma}_3 -\hat{\sigma}_3^2\hat{\rho}_{\textrm{\relsize{-2}{S}}}(t)-\hat{\rho}_{\textrm{\relsize{-2}{S}}}(t)\,\hat{\sigma}_3^2\right),
\end{eqnarray}
respectively, where $\hat{\sigma}_i^2=\hat{I}$ ($i=1,2,3$). These equations of motion, which do not have a Hamiltonian contribution, were written to emphasize their Lindblad form.

The Kraus representation is generally used in the area of quantum information processing to describe noisy dynamics. In particular, the operator-sum representation (\ref{Kraus}) of the above-defined channels is given by the Kraus operators, %respectively, %(0 y 1 estan OK???)
\begin{eqnarray}\label{eq:depolarizingK}
\hat{E}_0^{\textrm{\relsize{-2}{(D)}}}=\sqrt{1-\frac{3p}{4}}\hat{I}, \quad \hat{E}_{i}^{\textrm{\relsize{-2}{(D)}}}=\frac{\sqrt{p}}{2}\hat{\sigma}_{i}, \quad (i=1,2,3)\\
\label{eq:amplitudedampK}
\hat{E}_0^{\textrm{\relsize{-2}{(AD)}}} = \ket{0}\bra{0}+\sqrt{1-p}\ket{1}\bra{1},\quad \hat{E}_1^{\textrm{\relsize{-2}{(AD)}}} =\sqrt{p}\ket{0}\bra{1},\\
\label{eq:phasedampK}
\hat{E}_0^{\textrm{\relsize{-2}{(PD)}}} = \sqrt{1-p}\hat{I},\quad \hat{E}_1^{\textrm{\relsize{-2}{(PD)}}} = \sqrt{p} \ket{0}\bra{0},\quad \hat{E}_2^{\textrm{\relsize{-2}{(PD)}}} = \sqrt{p} \ket{1}\bra{1},
\end{eqnarray}
where $p(t)=1-e^{-\gamma\,t}$.  The superscripts indicate the case we are considering: depolarizing (D), amplitude-damping (AD) and phase-damping (PD) channels. A given physical process can be described by multiple Kraus representations, as can be seen in the phase-damping channel (\ref{eq:phasedampK}), which can also be given by the two Kraus operators \cite{Pirandola2008PRA77.032309}
\begin{equation}
 \hat{E}_0 = \sqrt{1-\frac{p}{2}}\hat{I}, \qquad \hat{E}_1=\sqrt{\frac{p}{2}}\hat{\gs}_3.
\end{equation} 

Traditionally, the term non-Markovian master equations denotes non-local equations of the form $d\hat{\rho}_{\textrm{\relsize{-2}{S}}}(t)/dt=\int_0^t d\tau \mathcal{K}(t-\tau)\hat{\rho}_{\textrm{\relsize{-2}{S}}}(\tau)$. However, not only they can be rewritten in time-local form $d\hat{\rho}_{\textrm{\relsize{-2}{S}}}(t)/dt= \mathcal{L}(t)\hat{\rho}_{\textrm{\relsize{-2}{S}}}(t)$ \cite{Andersson2007, Chruscinski2010}, but also as generalized Lindblad equations
\begin{eqnarray}
\label{eq:nonMarkov}
 \frac{d\hat{\rho}_{\textrm{\relsize{-2}{S}}}(t)}{dt}&=-\frac{i}{\hbar}\left[\hat{H}(t),\,\hat{\rho}_{\textrm{\relsize{-2}{S}}}(t)\right]\\
+&\hspace{-0.7cm}\sum_i\gamma_i(t) \left(2\,\hat{L}_i(t)\,\hat{\rho}_{\textrm{\relsize{-2}{S}}}(t)\,\hat{L}_i^\dagger(t)-\hat{L}_i^\dagger(t)\,\hat{L}_i(t)\,\hat{\rho}_{\textrm{\relsize{-2}{S}}}(t)-\hat{\rho}_{\textrm{\relsize{-2}{S}}}(t)\,\hat{L}_i^\dagger(t)\,\hat{L}_i(t)\right), \nonumber
\end{eqnarray} 
with time-dependent Lindblad operators $\hat{L}_i(t)$ and decay rates $\gamma_i(t)$ \cite{Breuer2004,Andersson2010}. The master equation (\ref{eq:nonMarkov}), where the decay rates become negative for some interval of time, is non-Markovian. In the following sections we show that the same Kraus representations used to described the usual one-qubit noisy channels can be related to many master equations: Markovian equations with constant decay rates, Markovian equations with time-dependent decay rates, and non-Markovian equations.

\section{Simple quantum microscopic model of a depolarizing channel}\label{sec:microdepolarizing}
Electronic spins in semiconductor quantum dots have been proposed as a prototype of a scalable quantum computer \cite{Loss1998PRA57.120}. A Caldeira-Leggett microscopic model of decoherence, considered to be a generic phenomenological description of the environment of the spins, has also been advanced \cite{CLM}. The interaction of the spin with the collection of harmonic oscillators, which describe its environment, is of the form
$
 \hat{H}_{\textrm{\relsize{-2}{int}}} = \lambda \sum_i \hat{\sigma}_i \sum_k g_k \left( \hat{a}_{ik}+\hat{a}_{ik}^\dag\right),
$
where $\hat{a}_{ik}$ ($\hat{a}_{ik}^\dag$) is the annihilation (creation) operator of the $k$-th harmonic oscillator coupled to the $i$-th component of the spin. The interaction is clearly isotropic. It is also natural to represent the degrees of freedom of the environment as spins, like in spin star models $\hat{H}_{\textrm{\relsize{-2}{ss}}} =  \frac{g}{\hbar} \hat{\boldsymbol{s}}\cdot \sum_{i} \hat{\boldsymbol{s}}_i$ \cite{Hutton2004PhysRevA.69.042312,Hamdouni2007PRB76.174306,Melikidze2004PRB70.014435}, where the interaction corresponds to the hyperfine interaction, or to the interaction of the electronic spins with other electronic spins. Repeated application of sum of environmental spins leads to a Hilbert space which is a direct sum of ``effective'' spins. Here, a model closedly related but generally different to the spin star model is considered. The model is described by the Hamiltonian
\begin{equation}
\label{eq:Hmudepola}
 \hat{H} = \oplus_k \frac{g_k}{\hbar} \hat{\boldsymbol{s}}\cdot \hat{\boldsymbol{S}}_{k}.
\end{equation}
It is important to pay attention to the direct sum of effective spins appearing in the Hamiltonian (\ref{eq:Hmudepola}). 

The model given by (\ref{eq:Hmudepola}) corresponds to the spin star model with identical coupling constants if the following conditions are obeyed. First, $g_k=g$; second, the dimension of the environmental Hilbert space is $D=2^N$; and third, the degeneracy of the angular momentum $l$, the number of values of $k$ with $l_k=l$, is \cite{Hutton2004PhysRevA.69.042312,Hamdouni2007PRB76.174306,Melikidze2004PRB70.014435}
\begin{equation}
\label{eq:degeneracy}
 \nu(N,l) = \left(\begin{array}{c}  N\\ \frac{N}{2}-l \end{array}\right)
 - \left(\begin{array}{c}   N\\ \frac{N}{2}-l-1  \end{array}\right),
\end{equation} 
Here, we have taken into account that the eigenvalue of $\hat{\boldsymbol{S}}_{k}^2$ is $\hbar^2 l_k(l_k+1)$.

The initial state of the total system is assumed to be factorized $\hat{\rho}_{\textrm{\relsize{-2}{T}}}(0)=\hat{\rho}_{\textrm{\relsize{-2}{S}}}(0)\otimes\hat{\rho}_{\textrm{\relsize{-2}{E}}}(0)$, that is, equal to the product of the initial spin state $\hat{\rho}_{\textrm{\relsize{-2}{S}}}(0)$ and the initial bath state, $\hat{\rho}_{\textrm{\relsize{-2}{E}}}(0)$. The latter is assumed to be the maximally mixed state $D^{-1} \hat{I}_{D\times D}$, where $\hat{I}_{D\times D}$ stands for the identity in $D$ dimensions, where $D=\sum_k d_k$ and $d_k$ is the dimension of the Hilbert state of the $k$-th environmental effective spin. Since the Hamiltonian is given by a direct sum, the dynamics corresponding to different values of $k$ are decoupled. In other words, we have many independent problems describing the interaction between spin $1/2$ and spin $l_k$, where the initial state is  $\hat{\rho}_{\textrm{\relsize{-2}{S}}}(0)\otimes (d_k)^{-1} \hat{{I}}_{d_k\times d_k}$, $d_k=(2l_k+1)$, with probability $q_k= d_
k/D$. 

We first solve the dynamical problem for a fixed value of $k$. It is convenient to make a unitary transformation, from the separated basis $\ket{s,s_{mk}}$ to the coupled basis $\ket{j_k,s_{ik}}$. States $\ket{s,s_{mk}}$, where $s=0,1$ and $s_{mk}=0,1,\cdots,2l_k$, are simultaneous eigenstates of $\hat{s}^2$, $\hat{S}_{k}^2$, $\hat{s}_z$ and $\hat{S}_{kz}$ with eigenvalues $3\hbar^2/4$, $\hbar^2 l_k (l_k+1)$, $\hbar(1/2-s)$ and $\hbar(l_k-s_{mk})$, respectively. Coupled basis states $\ket{j_k,s_{ik}}$ are simultaneous eigenstates of $\hat{s}^2$, $\hat{S}_{k}^2$, $\hat{J}_k^2$ and $\hat{J}_{kz}$, where $\hat{\boldsymbol{J}}_k = \hat{\boldsymbol{s}} +\hat{\boldsymbol{S}}_{k}$, with eigenvalues $3\hbar^2/4$, $\hbar^2 l_k (l_k+1)$, $\hbar^2j_k(j_k+1)$ and $\hbar(j_k-s_{ik})$, respectively. Here, $j_k$ can only take two values: $l_k+1/2$ and $l_k-1/2$, and $s_{ik}$ varies between zero and $2l_k+2$ ($2l_k$) in the former (latter) case. One can go from the separated basis to the coupled one using
\begin{equation}
\fl
 \left(\begin{array}{c} \Ket{l_k+\frac{1}{2}, s_{ik}}\\ \Ket{l_k-\frac{1}{2}, s_{ik}-1}\end{array}\right)
=\left(\begin{array}{lr} \cos\alpha_{ki} & \sin\alpha_{ki}\\-\sin\alpha_{ki} &\cos\alpha_{ki} \end{array}\right)
 \left(\begin{array}{c} \Ket{0,s_{ik}}\\ \Ket{1,s_{ik}-1} \end{array}\right),
\end{equation}
where the elements of the transition matrix are given by the Clebsh-Gordan coefficients
\begin{equation*}
 \sin\alpha_{ki} = \sqrt{\frac{s_{ik}}{2l_k+1}}, \qquad \cos\alpha_{ki} =\sqrt{\frac{2l_k-s_{ik}+1}{2l_k+1}},
\end{equation*}
and $s_{ik}=1,\cdots,2l_k$. The extreme cases, $s_{ik}=0$ and $s_{ik}=2l_k+1$, simply read $\Ket{l_k+\frac{1}{2}, s_{ik}=0} =\Ket{0,s_{ik}=0}$ and $\Ket{l_k+\frac{1}{2}, s_{ik}=2l_k+1} =\Ket{1,s_{ik}=2l_k}$.
Taking into account that the system's initial state is assumed to be pure $\hat{\rho}_{\textrm{\relsize{-2}{S}}}(0)=\ket{\psi_0}\bra{\psi_0}$, and that in the sector of the $k$-th effective environment spin,
$\hat{I}_{d_k}=\sum_{s_{mk}=0}^{2l_k} \ket{s_{mk}}\bra{s_{mk}}$, the dynamical problem with initial state $\hat{\rho}_{\textrm{\relsize{-2}{S}}}(0)\otimes (d_k)^{-1} \hat{{I}}_{d_k\times d_k}$ can be solved as the incoherent sum, with weights $(d_k)^{-1}$, of the problems with initial condition
\begin{equation*}
 \ket{\psi_{mk}(0)} =\ket{\psi_0}\otimes \ket{s_{mk}} =\left(c_0\ket{0}+c_1\ket{1}\right)\otimes \ket{s_{mk}} ,
\end{equation*}
where $s_{mk}=0,1,\cdots,2l_k$.
Since the spin-bath Hamiltonian can be recast as $\oplus_k{g_k} (\hat{J}_k^2-\hat{s}^2-\hat{S}_k^2)/({2\hbar})$, we can see that the coupled states $\ket{l_k+\frac{1}{2},s_{ik}}$ ($\ket{l_k-\frac{1}{2},s_{ik}}$) are eigenstates of $\hat{H}$ with eigenvalue $\hbar g_k l_k/2$ ($-\hbar g_k (l_k+1)/2$).  
If the initial state $\ket{\psi_{mk}(0)}$ is expanded in the coupled basis, evolved in time in this basis (in which the Hamiltonian is diagonal), and transformed back to the separated basis we obtain
\begin{eqnarray} \nonumber
\ket{\psi_{mk}(t)}=& c_0(t) \ket{0,s_{mk}} +c_1(t) \ket{1,s_{mk}} \\ &+c_2(t) \ket{0,s_{mk}+1} +c_3(t) \ket{1,s_{mk}-1}.
\end{eqnarray} 
The time-dependent coefficients are given by $c_a(t)=\exp(-igt/4)\tilde{c}_a(t),\, a=0,1,2,3$, where
% \begin{subequations}
\begin{eqnarray}
\tilde{c}_0(t)& =c_0\left(\cos\theta_t +i\frac{2s_{ik}-2l_k-1}{2l_k+1} \sin\theta_t\right),\\
\tilde{c}_1(t)& =c_1\left(\cos\theta_t +i\frac{2l_k-2s_{ik}-1}{2l_k+1} \sin\theta_t\right),\\
\tilde{c}_2(t)& =-2ic_1 \frac{\sqrt{(s_{ik}+1)(2l_k-s_{ik})}}{2l_k+1} \sin\theta_t,\\
\tilde{c}_3(t)& =-2ic_0 \frac{\sqrt{s_{ik}(2l_k-s_{ik}+1)}}{2l_k+1} \sin\theta_t,
\end{eqnarray}
% \end{subequations}
and $\theta_t=g_k(2l_k+1)t/4$.
After finding the total density matrix operator, and tracing out the bath degrees of freedom, the reduced density operator for the spin one-half system, $\hat{\rho}_{\textrm{\relsize{-2}{S}}mk}(t)=\tr_k(\ket{\psi_{mk}(t)}\bra{\psi_{mk}(t)})$, is found to be 
\begin{eqnarray*} 
 \hat{\rho}_{\textrm{\relsize{-2}{S}}mk}(t) =& \left(|c_2(t)|^2+|c_0(t)|^2\right)\Ket{0}\Bra{0} + c_0(t)\overline{c_1(t)}\Ket{0}\Bra{1} \\ 
&+ c_1(t)\overline{c_0(t)} \Ket{1}\Bra{0} + \left(|c_1(t)|^2+|c_3(t)|^2\right) \Ket{1}\Bra{1},
\end{eqnarray*}
where $\overline{c}$ is the complex conjugate of $c$. 
Now, it is possible to take into account that the initial bath spin-$k$ state is $(d_k)^{-1}\hat{I}_{d_k\times d_k}$, to show that 
\begin{equation}
 \hat{\rho}_{\textrm{\relsize{-2}{S}}k}(t) = \frac{1}{2l_k+1} \sum_{s_{mk}=0}^{2l_k} \hat{\rho}_{\textrm{\relsize{-2}{S}}mk}(t) =\frac{1}{2}\left(\hat{I}_{2\times 2}+\boldsymbol{s}_k(t)\cdot\hat{\boldsymbol{\sigma}}\right),
\end{equation} 
can be exactly calculated. The Bloch vector $\boldsymbol{s}_k(t)$ completely characterizes the quantum state of spin one-half system and is given by
\begin{equation}
 \boldsymbol{s}_k(t)= \boldsymbol{s}(0) \frac{4 l_k^2+4 l_k+3+8l_k (l_k+1) \cos \left(\frac{(2l_k+1)g_kt}{2}\right)}{3 (2 l_k+1)^2},
\end{equation} 
where $\boldsymbol{s}_k(0)$ is the value of the Bloch vector of the initial state of the system. 
For large values of $l_k$, $\boldsymbol{s}_k(t)\approx \boldsymbol{s}(0)(1+2\cos \left({(2l_k+1)g_kt}/{2}\right))/3$. 
Finally, we have to average over the different values of $k$: $\hat{\rho}_{\textrm{\relsize{-2}{S}}}(t)=\sum_k q_k \hat{\rho}_{\textrm{\relsize{-2}{S}}k}(t)$. The final form of the Bloch vector is
\begin{equation}
 \boldsymbol{s}(t)= \boldsymbol{s}(0) \sum_k q_k \frac{4 l_k^2+4 l_k+3+8l_k (l_k+1) \cos \left(\frac{(2l_k+1)g_kt}{2}\right)}{3 (2 l_k+1)^2}.
\end{equation} 

In the continuum limit, assuming a fixed value of $l_k=l$ and a gaussian distribution of coupling constants, the final expression for the Bloch vector is
\begin{equation}
\label{eq:depol1}
 \boldsymbol{s}(t)= \boldsymbol{s}(0) \frac{4 l^2+4 l+3+8l (l+1) e^{-(2l+1)^2 \gs^2 t^2/8}\cos \left(\frac{(2l+1)Gt}{2}\right)}{3 (2 l+1)^2},
\end{equation} 
where $\gs$ is the standard deviation. The probability density to have a particular value $g$ for the coupling constant has been assumed to be
\begin{equation}
\label{eq:gaussian}
q_G(g) = (2\pi\gs^2)^{-1/2}\exp(- (g-G)^2/(2\gs^2)),
\end{equation} 
where $q_G(g)$ is the continuous version of $q_k$.
Note that the Bloch vector decreases until its length is reduced for a factor $x(l)=\frac{4 l^2+4 l+3}{3 (2 l+1)^2}$, which satisfies the inequality $\frac{1}{3}< x(l)\leq \frac{1}{2}$. 

If the depolarizing channel (\ref{eq:depolarizingK}) is applied to an initial pure state, its Bloch's vector evolves as $\boldsymbol{s}^{\textrm{\relsize{-2}{(D)}}}(t)=\boldsymbol{s}^{\textrm{\relsize{-2}{(D)}}}(0) (1-p)$. We thereby conclude that we have a depolarizing channel for which 
\begin{equation}
\label{eq:nonmarkdep1}
 p(t) = \frac{8l(l+1)}{3(2l+1)^2}\left[1-e^{-(2l+1)^2 \gs^2 t^2/8}\cos \left(\frac{(2l+1)Gt}{2}\right)\right].
\end{equation} 
We first consider the case $G=0$. We have a gaussian decay very different from the (typical) Markovian process, $p(t)=1-e^{-\gamma t}$. Moreover, in contrast with the (typical) Markovian case, where $p(t)$ varies from zero to one, in the process described by (\ref{eq:depol1}), $p(t)$ varies, monotonically, from zero to $p_{\inf}(l)= \frac{8l(l+1)}{3(2l+1)^2}<\frac{2}{3}$, i.e., this process does not display complete depolarization. However, this process is Markovian in the less restricted sense outlined at the end of section \ref{sec:MLK}. Indeed, $\gamma(t)=-\dot{p}(t)/(1-p(t))$, where the dot stands for the time derivative, is a strictly non-negative function of time. For any non-vanishing value of $G$ true non-Markovian effects are present. However, they can be very small, unless $G$ is at least of the order of $\sigma$. In this case, $\gamma(t)$ becomes temporarily negative. 

As a second example of the distribution of coupling constants, we assume a Lorentzian distribution $q(g) =a(\pi(g^2+a^2))^{-1}$, where $a$ is the scale parameter which specifies the half-width at half-maximum. The average Bloch vector decreases as in the previous example
\begin{equation}
\label{eq:depol2}
 \boldsymbol{s}_l(t)= \boldsymbol{s}_l(0) \frac{4 l^2+4 l+3+8l (l+1) e^{-(2l+1) a t/2}}{3 (2 l+1)^2},
\end{equation} 
but in an exponential way. The corresponding master equation reads
\begin{equation}
\label{eq:nonmarkdep3}
\frac{d\hat{\rho}_{\textrm{\relsize{-2}{S}}}(t)}{dt}=\frac{\gamma(t)}{8}\sum_{i=1}^3\left(2\hat{\sigma}_i \hat{\rho}_{\textrm{\relsize{-2}{S}}}(t) \hat{\sigma}_i-\hat{\sigma}_i^2\hat{\rho}_{\textrm{\relsize{-2}{S}}}(t)-\hat{\rho}_{\textrm{\relsize{-2}{S}}}(t)\hat{\gs}_i^2\right), 
\end{equation}
where $\hat{\sigma}_i^2=\hat{I}$, which displays a generalized Lindblad form with $\gamma(t)$ an explicit function of time
\begin{equation}
\label{eq:firstgammat}
 \gamma(t)=\Gamma\left(1-\frac{4 l^2+4 l+3}{4 l^2+4 l+3+8l (l+1) e^{-(2l+1) a t/2}} \right).
\end{equation} 
This process is Markovian, in the sense that $\gamma(t)$ is non-negative.

In a simpler, but also interesting case, $l$ and $g$ fixed, $p(t)$ varies periodically, and the master equation is given by (\ref{eq:nonmarkdep3}) with a time-dependent  $\gamma(t)$,
\begin{equation}
\label{eq:secondgammat}
 \gamma(t)=\frac{(2l+1)g}{2}\frac{8l (l+1)\sin\left(\frac{(2l+1)gt}{2} \right)}{4 l^2+4 l+3+8l (l+1) \cos\left(\frac{(2l+1)gt}{2} \right)} ,
\end{equation} 
which, in contrast to (\ref{eq:firstgammat}), attain negative values, and characterizes a true non-Markovian process. This case is an extreme case of gaussian distribution of coupling constants, where the dispersion $\sigma$ goes to zero. Uniform distributions of the coupling constant, between a lower and an upper limit, and the spin star model with identical coupling constants also produce non-Markovian evolution. 

\section{Simple classical microscopic model of a depolarizing channel}\label{sec:macrodepolarizing}

\begin{figure}[h!]
\centering
\includegraphics[height=7cm]{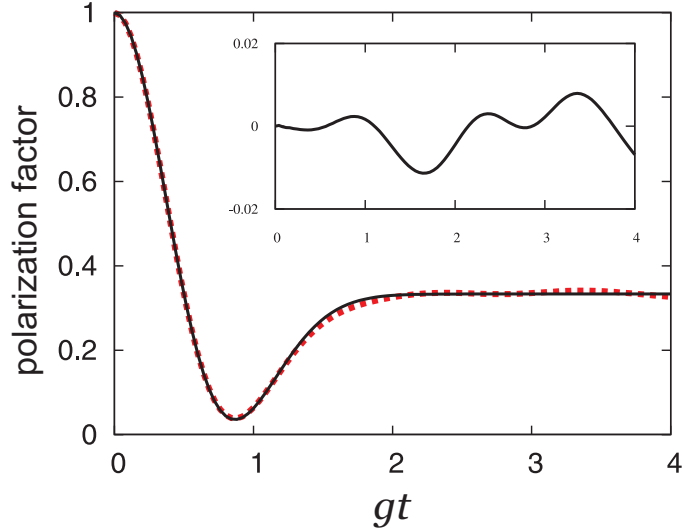}
\caption{\footnotesize Polarization factor $f(t)$ for a microscopic model with classical gaussian noise, with zero average and unit standard deviation. The solid line represents the theory and the dashed red line a numerical average over 10000 realizations. In the inset, the difference between the numerical result and the theoretical prediction is plotted as a function of $gt$.}
\label{fig:foft}
\end{figure}

In nuclear magnetic resonance experiments and Bose-Einstein condensates the decoherence process is often caused by residual fluctuating magnetic fields (see, for example, \cite{Specht2011Nature473.190}), which can be considered classical. In the Hamiltonian description of this process
\begin{equation}
 \hat{H}= -\hat{\boldsymbol{\mu}}\cdot \boldsymbol{B}=\hbar g \boldsymbol{\xi}\cdot \hat{\boldsymbol{\sigma}},
\end{equation} 
the dimensionless independent random variables $\xi_i$, $i=1,2,3$, assumed to be gaussian of zero average and standard deviation $\gs$, are proportional to the corresponding magnetic field components. The constant $g$ have units of frequency. If the initial state of the spin system is $\hat{\rho}_{\textrm{\relsize{-2}{S}}}(0)$, its state at time $t$ is
\begin{equation}
\hat{\rho}_{\textrm{\relsize{-2}{S}}}(t) = e^{-igt\boldsymbol{\xi}\cdot \hat{\boldsymbol{\sigma}}}\hat{\rho}_{\textrm{\relsize{-2}{S}}}(0)e^{igt\boldsymbol{\xi}\cdot \hat{\boldsymbol{\sigma}}}
= \frac{1}{2}\left(\hat{I}_{2\times 2}+\boldsymbol{s}(t)\cdot\hat{\boldsymbol{\sigma}}\right),
\end{equation} 
where
\begin{equation}
\boldsymbol{s}(t) = \boldsymbol{s}(0)\cos(2gt\xi)+\boldsymbol{\xi}\times\boldsymbol{s}(0)\frac{\sin(2gt\xi)}{\xi}+(\boldsymbol{\xi}\cdot\boldsymbol{s}(0))\frac{\boldsymbol{\xi}(1-\cos(2gt\xi))}{\xi^2},
\end{equation} 
and $\xi=\sqrt{\boldsymbol{\xi}\cdot\boldsymbol{\xi}}$. Averaging over the different realizations of the noise variables $\xi_i$ ($i=1,2,3$), the state of the two-level system is
\begin{eqnarray}
\hat{\overline{\rho}}_{\textrm{\relsize{-2}{S}}}(t) &=&\frac{1}{2}\left(\hat{I}_{2\times 2}+\overline{\boldsymbol{s}}(t)\cdot\hat{\boldsymbol{\sigma}}\right),
\nonumber\\
\bar{s}_i(t)& =& s_i(0)\left(\overline{\cos(gt\xi)}+\overline{\frac{\xi_i^2(1-\cos(gt\xi))}{\xi^2}}\right)= s_i(0)f(t),
\end{eqnarray} 
where the property of vanishing averages of the distributions of $\xi_i$ were used. Employing the gaussian probability distributions $q_G(\xi_i)= (2\pi\gs^2)^{-1/2}\exp(- \xi_i^2/(2\gs^2))$ ($i=1,2,3,$) and transforming to spherical coordinates, one finds the polarization factor 
\begin{equation}
 f(t) = \frac{1}{3}\left( 1+2(1-4g^2\sigma^2 t^2)e^{-2g^2\sigma^2 t^2}\right),
\end{equation}
that is directly connected to the parameter $p(t)$ of the depolarizing channel by $p(t)=1-f(t)$. 
The spherical coordinates were defined by $r=\xi$, $\theta=\arccos(\xi_3/\xi)$, and $\phi=\arctan(\xi_2/\xi_1)$, where $r\in[0,\infty)$, $\phi\in [0,2\pi)$ and $\theta\in [0,\pi]$. 
In Figure \ref{fig:foft} we plot the polarization factor for $\sigma=1$ as a function of the dimensionless time $gt$. In this case, $\gamma(t)=-\dot{f}(t)/f(t)$ also attains the negative values that characterize non-Markovian processes.

We stress that albeit none of the examples considered in the previous section and this section are Markovian in the restricted sense, all of them are described by the Kraus representation (\ref{eq:depolarizingK}) of the depolarization channel, with different definitions of the parameter $p$. The time-dependence of $p$ allows for non-exponential and partial decay, in time-dependent Markovian processes, and even recoherence in non-Markovian process, in which case $\gamma(t)$ is negative.

\section{Simple quantum microscopic model of a dephasing channel}\label{sec:microdephasing}

The model Hamiltonian we consider, given by 
\begin{equation} \label{uno}
\hat{H} = \frac{\hbar\omega(t)}{2}\hat{\sigma}_3 +\sum_k\hbar\omega_k \left(\hat{a}_k+\frac{c_k\hat{\sigma}_3}{\omega_k}\right)^\dag \left(\hat{a}_k+\frac{c_k\hat{\sigma}_3}{\omega_k}\right),
\end{equation}
describes the interaction of a two-level system with a collection of oscillators (coupling constants $c_k$), through an interaction which conserves the system's observable $\hat{\sigma}_3$. The creation and annihilation operators $\hat{a}_k^\dagger$ and $\hat{a}_k$ satisfy the usual boson commutation relations $[\hat{a}_k,\hat{a}_{k^\prime}^\dagger]=\delta_{kk^\prime}$. Hamiltonians of the type we consider here have been explored by many authors \cite{Collet1988PRA38.2233,Luczka1990PhysA167.919,Kampen1995JStatPhys78.299,Palma1996RoySocLonProcSA452.567,FonsecaRomero1997PLA235.432}.  We assume an initial state of the Feynman-Vernon form $\hat{\rho}_{\textrm{\relsize{-2}{T}}}(0)=\hat{\rho}_{\textrm{\relsize{-2}{S}}}(0)\otimes \prod_k \hat{\rho}_k(0)$, where $\hat{\rho}_k(0) = \exp\left(-\beta\hbar\omega_k\hat{a}_k^\dag \hat{a}_k\right)/ Z_k$ is the thermal state of the $k$-th mode at inverse temperature $\beta=1/(k_B T)$ and $Z_k=$tr$\left(\exp\left(-\beta\hbar\omega_k\hat{a}_k^\dag \hat{a}_k\right)\right)$. If 
the action of the unperturbed Hamiltonians is 
separated from the interaction, the 
total density operator is
\begin{equation}
\fl
\hat{\rho}_{\textrm{\relsize{-2}{T}}}(t) =e^{-i\Omega(t)\hat{\sigma}_3/2} e^{-i\sum_k\omega_k t \hat{a}_k^\dag \hat{a}} \mathcal{U}_I(t,0)
\hat{\rho}_{\textrm{\relsize{-2}{T}}}(0) \mathcal{U}_I^\dag(t,0)
e^{i\sum_k\omega_kt\hat{a}_k^\dag \hat{a}}e^{i\Omega(t)\hat{\sigma}_3/2}
\end{equation} 
where the function $\Omega(t)=\int_0^td\tau \omega(\tau)$ has been defined. The interaction evolution operator in the interaction picture, $\mathcal{U}_I(t,0)$,
\begin{equation}
\fl
 \mathcal{U}_I(t,0) =\mathcal{T}\exp\left(-i\int_0^t d\tau \sum_k \left(c_k e^{i\omega_k \tau}\hat{a}_k^\dag \hat{\sigma}_3+h.c.\right)\right)=
\mathcal{T} e^{-i\int_0^t d\tau \sum_k \hat{H}_k(\tau)}
\end{equation} 
needs the time-ordering prescription, indicated by $\mathcal{T}$, because the commutators $[\hat{H}_k(\tau),\hat{H}_k(\tau^\prime)]$ do not vanish. However, due to the simplicity of the Lie algebra satisfied by the operators appearing in the interaction Hamiltonian, $[\hat{a}_k \hat{\sigma}_3, \hat{a}_l^\dag \hat{\sigma}_3]=\delta_{k,l}$, it is relatively easy to trace out the environmental degrees of freedom to write
\begin{equation}
\fl
 \hat{\rho}_{\textrm{\relsize{-2}{S}}}(t) =e^{-i\Omega(t)\hat{\sigma}_3/2} \Tr_k\left(e^{-i\int_0^t d\tau \sum_k \hat{H}_k(\tau)}
\hat{\rho}_{\textrm{\relsize{-2}{S}}}(0)\prod_k\hat{\rho}_k(0) e^{-i\int_0^t d\tau \sum_k \hat{H}_k(\tau)}
\right)
e^{i\Omega(t)\hat{\sigma}_3/2}.
\end{equation} 
Employing algebraic techniques \cite{Louisell1973Q} to calculate the trace, the density operator of the system can be written as 
\begin{equation}
\label{eq:rhoPD1}
 \hat{\rho}_{\textrm{\relsize{-2}{S}}}(t) = e^{-i\Omega(t)\hat{\sigma}_3/2}\left(e^{-\Gamma(t)(1-\hat{\sigma}_3\bullet\hat{\sigma}_3)/2}\hat{\rho}_{\textrm{\relsize{-2}{S}}}(0)\right)e^{i\Omega(t)\hat{\sigma}_3/2},
\end{equation} 
where  $4\Gamma(t)=\sum_k |c_k|^2(1-\cos(\omega_k t))\coth(\hbar\omega_k\beta/2)/\omega_k^2$. Here we have used the dot superoperator convention, where the dot ``$\bullet$'' stands for the operator to the right of the superoperator as illustrated below
\begin{equation*}
 e^{x\hat{\sigma}_3\bullet \hat{\sigma}_3}\hat{\rho}_{\textrm{\relsize{-2}{S}}}(0)=
\sum_{n=0}^\infty \frac{x^n}{n!}\left(\hat{\sigma}_3\bullet \hat{\sigma}_3\right)^n\hat{\rho}_{\textrm{\relsize{-2}{S}}}(0)=
\sum_{n=0}^\infty \frac{x^n}{n!}\hat{\sigma}_3^n\hat{\rho}_{\textrm{\relsize{-2}{S}}}(0)\hat{\sigma}_3^n.
\end{equation*} 
Taking into account that $\hat{\rho}_{\textrm{\relsize{-2}{S}}}(0)=\sum_{ij}\rho_{ij}\ket{i}\bra{j}$ the expression into brackets in equation (\ref{eq:rhoPD1}) can be simplified as follows
\begin{eqnarray}
 e^{-\Gamma(t)(1-\hat{\sigma}_3\bullet\hat{\sigma}_3)/2}\hat{\rho}_{\textrm{\relsize{-2}{S}}}(0)&=&e^{-\Gamma(t)(1-\hat{\sigma}_3\bullet\hat{\sigma}_3)/2}\sum_{ij}\rho_{ij}\ket{i}\bra{j}\nonumber\\
&=&\sum_{ij}\rho_{ij}e^{-\Gamma(t)(1-s_z(i)s_z(j))/2}\ket{i}\bra{j},
\end{eqnarray} 
with $s_z(0)=1=-s_z(1)$. If we explicitly write the four terms we have
\begin{equation*}
\fl
 e^{-\Gamma(t)(1-\hat{\sigma}_3\bullet\hat{\sigma}_3)/2}\hat{\rho}_{\textrm{\relsize{-2}{S}}}(0) = \rho_{00}\ket{0}\bra{0} +e^{-\Gamma(t)}\rho_{01}\ket{0}\bra{1}  +e^{-\Gamma(t)}\rho_{10}\ket{1}\bra{0} +\rho_{11}\ket{1}\bra{1},
\end{equation*} 
which can be recast as
\begin{equation}
\fl
 e^{-\Gamma(t)(1-\hat{\sigma}_3\bullet\hat{\sigma}_3)/2}\hat{\rho}_{\textrm{\relsize{-2}{S}}}(0) = e^{-\Gamma(t)}\hat{\rho}_{\textrm{\relsize{-2}{S}}}(0)+(1- e^{-\Gamma(t)})\left(\rho_{00}\ket{0}\bra{0}+\rho_{11}\ket{1}\bra{1}\right).
\end{equation} 
We recognize the form of a phase damping channel with $p(t)=1-e^{-\Gamma(t)}$. Finally, including the effect of the unitary operators $ e^{\pm i\Omega(t)\hat{\sigma}_3/2}$, we see that, at time $t$, the reduced density matrix of the system can be written as a phase damping (PD) channel, with $\widetilde{E}_i^{\textrm{\relsize{-2}{(PD)}}}=e^{-i\Omega(t) \hat{\sigma}_3/2} \hat{E}_i^{\textrm{\relsize{-2}{(PD)}}}$, where the Kraus operators $\hat{E}_i^{\textrm{\relsize{-2}{(PD)}}}$ ($i=0,1,2$) are given in Eq. (\ref{eq:phasedampK}). 

\begin{figure}[h!]
\centering
\includegraphics[height=7cm]{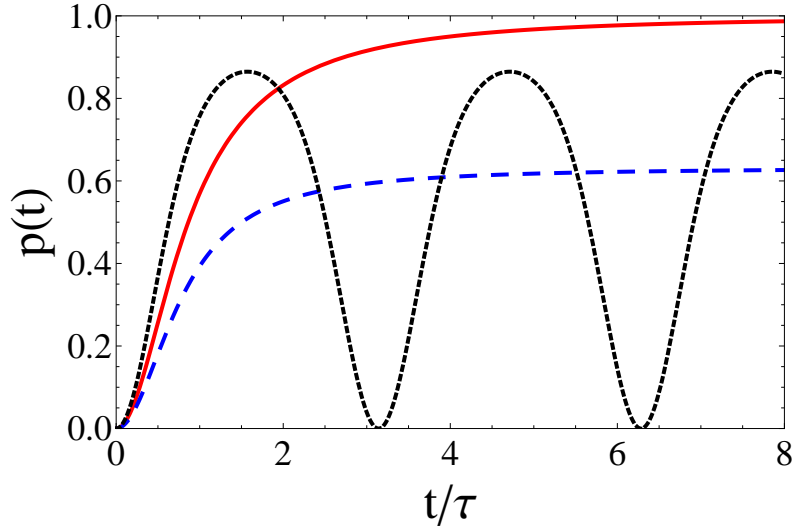}
%Dephasing_Lambda.eps: 0x0 pixel, 300dpi, 0.00x0.00 cm, bb=0 0 908 573
\caption{\footnotesize Dephasing channel. Function $p(t)$ for a single oscillator with frequency $1/(2\tau)$ and $|c|^2\coth(\hbar\omega\beta/2)/\omega^2=4$ (dotted line), and for a continuum of oscillators with $J(\omega)=8\pi\omega e^{-\omega \tau}$ at zero temperature (solid line) and at $\beta=\tau/\hbar$ (dashed line). Time is measured in units of the timescale $\tau$ ($1/\tau$ is a cut-off frequency).}
\label{fig:poft}
\end{figure}

The behavior of $p(t)$ depends on the number of oscillators of the environment, their frequencies, coupling constants and the temperature of the bath. If the ``environment'' contains a single oscillator, $p$ varies periodically, with the unperturbed frequency of this oscillator, between $0$ and $p_M$: the higher the temperature (or the stronger the coupling or the smaller the frequency) the greater the value of $p_M$. In this case, $p(t)$ does not have a limit for long times. Moreover, the master equation, which can be written as
\begin{equation}
\frac{d\hat{\rho}_{\textrm{\relsize{-2}{S}}}}{dt} = \frac{1}{i\hbar}[\hbar\omega(t)\hat{\sigma}_3,\hat{\rho}_{\textrm{\relsize{-2}{S}}}] +\frac{\gamma(t)}{4}\left(2\hat{\sigma}_3\hat{\rho}_{\textrm{\relsize{-2}{S}}}\hat{\sigma}_3 -\hat{\sigma}_3^2\hat{\rho}_{\textrm{\relsize{-2}{S}}}-\hat{\rho}_{\textrm{\relsize{-2}{S}}}\hat{\sigma}_3^2 \right),
\end{equation}  
where $\hat{\sigma}_3^2=\hat{I}$, is non-Markovian. 
In particular, $\gamma(t)$ is negative for an infinite number of time intervals. Differently, when the limit to the continuum is taken  ($\sum_k\rightarrow\int d\omega/(2\pi)$), in the decoherence function $\Gamma(t)$ at finite temperature a zero-temperature contribution can be isolated, 
\begin{eqnarray}
\label{lambdazero}
\Gamma_{0} (t) & = & \frac{1}{4}\int_0^{\infty} \frac{J(\omega)}{2\pi\omega} \left(1-\cos(\omega t)\right) d\omega.
\end{eqnarray}
The symbol $J(\omega)$ stands for spectral density, that is the square of the coupling constant times the density of states divided by the frequency.
The behavior of $\Gamma(t)$ often differs significantly at large times from that of $\Gamma_{0}(t)$,
as shown in Figure \ref{fig:poft}. The examples given before illustrate how $p(t)=1-e^{-\Gamma(t)}$ can oscillate, grow to one or to a constant smaller than one, exponentially or not.

\section{Simple classical microscopic model of a dephasing channel}\label{sec:macrodephasing}
A dephasing channel also occurs when the magnitude of a classical magnetic field changes randomly. The Hamiltonian which describes this situation, $\hat{H}=\hbar\omega(t)\hat{\sigma}_3/2+\hbar g \xi(t)\hat{\sigma}_3$, where $\xi(t)$ is a random variable. This Hamiltonian has previously been investigated as a model of a solid state system, under the assumption that the random variable is a stochastic process producing random telegraphic noise \cite{LoFranco}. The equation of motion for the density operator is easily integrated,
\begin{eqnarray*}
\fl
\hat{\rho}_{\textrm{\relsize{-2}{S}}}(t)  = e^{-i\Omega(t)[\hat{\sigma}_3,\bullet]}&&\left( e^{-i g[\int_0^t\gd \xi(\tau) \hat{\sigma}_3,\bullet]}\hat{\rho}_{\textrm{\relsize{-2}{S}}}(0) \right) \\
\fl
\hspace{0.76cm}=  e^{-i\Omega(t)[\hat{\sigma}_3,\bullet]} && \left(
\sum_{n=0}^\infty \frac{(-ig)^n}{n!} 
\!\int_0^t \! dt_1 \cdots \int_0^t \! dt_n \xi(t_1) \cdots \xi(t_n) [\hat{\sigma}_3,\bullet]^n \hat{\rho}_{\textrm{\relsize{-2}{S}}}(0) \right),
\end{eqnarray*}
where $\Omega(t)=\int_0^t \omega(\tau)d\tau$, and the dot superoperator convention has been used. A gaussian stationary process  $\xi(t)$  with zero average satisfies \cite{Risken1989a}
\begin{eqnarray}
\label{eq:sgp1}\overline{\xi(t_1) \xi(t_2)} & = \Phi(t_1-t_2),\\
\label{eq:sgp2}\overline{\xi(t_1) \cdots \xi(t_{2n-1})} & = 0,\\
\label{eq:sgp3}\overline{\xi(t_1) \cdots \xi(t_{2n})} & = \sum_{\textrm{\relsize{-2}{Perm}}} \Phi(t_{i1}-t_{i2}) \cdots \Phi(t_{i(2n-1)}-t_{i(2n)}),
\end{eqnarray}
where $n$ is a positive integer, the overline indicates the expected value, and $\Phi(t_1-t_2)$ denotes the correlation function. 
A gaussian stationary process is given by
\begin{equation}
\label{eq:sgp}
 \xi(t) = \sum_i x_i \cos(\omega_i t) +y_i \sin(\omega_i t),
\end{equation}
where $x_i$ and $y_i$ are independent gaussian random variables with zero mean and identical standard deviation $\sigma_i$. 
By employing the definition  (\ref{eq:sgp}), one can verify the properties (\ref{eq:sgp1}) to  (\ref{eq:sgp3}). In particular,
the correlation function is given by
\begin{equation}
 \Phi(t_1-t_2) = \sum_i \sigma_i^2 \cos(\omega_i (t_1-t_2)).
\end{equation} 

The averaged density operator
\begin{eqnarray*}
\hat{\overline{\rho}}_{\textrm{\relsize{-2}{S}}}(t) & = e^{-i\Omega(t)[\hat{\sigma}_3,\bullet]}\left(
\sum_{n=0}^\infty \frac{(-ig)^{2n}}{(2n)!} 
\!\int_0^t \! dt_1 \cdots \int_0^t \! dt_n \times \right.\qquad \qquad\\
&\left. \qquad\qquad \times \sum_{\textrm{\relsize{-2}{Perm}}}
\Phi(t_{i1}-t_{i2}) \cdots \Phi(t_{i(2n-1)}-t_{i(2n)}) [\hat{\sigma}_3,\bullet]^{2n} \hat{\rho}_{\textrm{\relsize{-2}{S}}}(0) \right)
\end{eqnarray*}
can be simplified taking into account that there are $(2n)!/(2^n n!)$ permutations of the $2n$ times $t_i$ 
\begin{equation*}
\fl
\hat{\overline{\rho}}_{\textrm{\relsize{-2}{S}}}(t)  = e^{-i\Omega(t)[\hat{\sigma}_3,\bullet]} \left(  
\sum_{n=0}^\infty \frac{(-ig)^{2n}}{(2n)!} \frac{(2n)!}{2^n n!}
\left(\int_0^t \!\! dt_1 \! \int_0^t \!\! dt_2 
\Phi(t_{1}-t_{2})\right)^n  [\hat{\sigma}_3,\bullet]^{2n} \hat{\rho}_{\textrm{\relsize{-2}{S}}}(0)\right).
\end{equation*}
The dynamics of the two-level system
\begin{equation*}
\hat{\overline{\rho}}_{\textrm{\relsize{-2}{S}}}(t)  = 
e^{-i\Omega(t)[\hat{\sigma}_3,\bullet]} \left( e^{- \frac{g^2}{2}\int_0^t \! dt_1 \int_0^t dt_2 
\Phi(t_{1}-t_{2}) [\hat{\sigma}_3,[\hat{\sigma}_3,\bullet]]}
\hat{\rho}_{\textrm{\relsize{-2}{S}}}(0)\right),
\end{equation*}
can also be written in the more explicit form
\begin{equation}
\hat{\overline{\rho}}_{\textrm{\relsize{-2}{S}}}(t)  = 
e^{-i\Omega(t)\hat{\sigma}_3} \left( e^{- \frac{1}{2}\Gamma(t) (1-\hat{\sigma}_3 \bullet \hat{\sigma}_3)}
\hat{\rho}_{\textrm{\relsize{-2}{S}}}(0)\right)
 e^{i\Omega(t)\hat{\sigma}_3},
\end{equation}
where $\Gamma(t) = 2 g^2 \int_0^t \! dt_1 \int_0^t dt_2 \Phi(t_{1}-t_{2})$. As shown in the previous section, this evolution corresponds to a phase damping channel with $p(t)=1-e^{-\Gamma(t)}$. If the two-point correlation function $\Phi(t_{1}-t_{2})=\sigma^2 \delta(t_1-t_2)$ the dynamics is Markovian in the restricted sense. If $d\Gamma(t)/dt$ becomes negative, the dynamics is non-Markovian.

\section{Simple quantum microscopic model of an amplitude damping channel}\label{sec:microdamping}

Finally, we consider a qubit interacting with a collection of harmonic oscillators, which models the degrees of freedom of its environment, described by the Hamiltonian $\hat{H} = \frac{\hbar\omega}{2}\ket{1}\bra{1}+\sum_k\hbar\omega_k \hat{a}_k^\dag \hat{a}_k+\sum_k \left( c_k \hat{a}_k^\dag\hat{\sigma}_- +c_k^* \hat{a}_k \hat{\sigma}_+\right)$ \cite{Garraway1997PRA55.2290,Breuer2002,Maniscalco2006PRA73.012111}. Here, $\hat{a}_k$ is the annihilation operator of the $k$-th mode of the environment of free frequency $\omega_k$, and $c_k$ are interaction constants. If the initial state of qubit and environment is $\ket{\psi_{\textrm{\relsize{-2}{S}}}}\otimes \prod_k \ket{0}_k$, where $\ket{\psi_{\textrm{\relsize{-2}{S}}}}=\alpha\ket{1}+\beta\ket{0}$ with $|\alpha|^2+|\beta|^2=1$, the state at time $t$ can be written as
\begin{equation}
\fl
 \ket{\Psi(t)} = \alpha(t)\ket{1}\otimes \prod_k \ket{0}_k +\beta(t)\ket{0}\otimes \prod_k \ket{0}_k +\sum_l\delta_l(t)\ket{0}\otimes \ket{1}_l \otimes \prod_{k\neq l} \ket{0}_k. 
\end{equation} 
The use of Schr\"odinger equation leads to a set of coupled first-order linear equations for the coefficients $\alpha(t), \beta(t)$ and $\delta_l(t)$. Solving the equations for $\delta_l(t)$ in terms of $\alpha(t)$, and replacing into the equation for $\alpha(t)$ we find the integro-differential equation
\begin{equation}
 \frac{d\alpha(t)}{dt}+i\omega\alpha(t)+\int_0^t d\tau\sum_l |c_l|^2 e^{-i\omega_l(t-\tau)}\alpha(\tau)=0.
\end{equation} 
The solution of this equation, which is also obtained for the independent oscillator model \cite{FonsecaRomero2003b}, can be written as $\alpha(t) = \alpha \exp(-\Gamma(t)/2-i\Omega(t))$, where $\Gamma(t)\geq 0$ and $\Omega(t)$ are real functions. The reduced density matrix of the qubit is obtained tracing out the state of the oscillators, $ \hat{\rho_{\textrm{\relsize{-2}{S}}}}(t) =\tr_{\textrm{\relsize{-2}{E}}} \ket{\Psi(t)}\bra{\Psi(t)}$, that gives
\begin{equation}
 \hat{\rho}_{\textrm{\relsize{-2}{S}}}(t) = \mu(t)\ket{1}\bra{1}+(1-\mu(t))\ket{0}\bra{0}+\left( \alpha(t)\beta^*\ket{1}\bra{0}+h.c.\right),
\end{equation} 
where $\mu(t)=|\alpha(t)|^2=|\alpha|^2\exp(-\Gamma(t))$. A unitary contribution to the dynamics of the qubit can be isolated
\begin{eqnarray}\label{evolvedstateAD}
 \hat{\rho_{\textrm{\relsize{-2}{S}}}}(t) &=& e^{-i\Omega(t)\ket{1}\bra{1}}\Big[\mu(t)\ket{1}\bra{1}+(1-\mu(t))\ket{0}\bra{0}\nonumber\\
 &&+\left(e^{-\Gamma(t)/2}\alpha\beta^*\ket{1}\bra{0}+h.c.\right) \Big]  e^{i\Omega(t)\ket{1}\bra{1}}.
\end{eqnarray} 
Now, applying the Kraus operators $\hat{E}_0^{\textrm{\relsize{-2}{(AD)}}}$, $\hat{E}_1^{\textrm{\relsize{-2}{(AD)}}}$ of the amplitude damping channel of Eq. (\ref{eq:amplitudedampK}) to the initial state $\hat{\rho_{\textrm{\relsize{-2}{S}}}}(0)=\ket{\psi_{\textrm{\relsize{-2}{S}}}}\bra{\psi_{\textrm{\relsize{-2}{S}}}}$, one obtains
\begin{eqnarray}
\hat{E}_0^{\textrm{\relsize{-2}{(AD)}}}\hat{\rho_{\textrm{\relsize{-2}{S}}}}(0)(\hat{E}_0^{\textrm{\relsize{-2}{(AD)}}})^\dagger
&=&(1-p)|\alpha|^2\ket{1}\bra{1}+|\beta|^2\ket{0}\bra{0}\nonumber\\&&+\left(\sqrt{1-p}\alpha\beta^*\ket{1}\bra{0}+h.c.\right),\nonumber\\
\hat{E}_1^{\textrm{\relsize{-2}{(AD)}}}\hat{\rho_{\textrm{\relsize{-2}{S}}}}(0)(\hat{E}_1^{\textrm{\relsize{-2}{(AD)}}})^\dagger &=& p|\alpha|^2 \ket{0}\bra{0}.
\end{eqnarray}
Comparing the Kraus representation, $\sum_{i=0}^{1}\hat{E}_i^{\textrm{\relsize{-2}{(AD)}}}(t) \hat{\rho_{\textrm{\relsize{-2}{S}}}}(0)  (\hat{E}_i^{\textrm{\relsize{-2}{(AD)}}})^\dagger(t)$, with the evolved state of Eq.~(\ref{evolvedstateAD}), we conclude that $p(t)=1-e^{-\Gamma(t)}$ and that the qubit dynamics corresponds to an
amplitude damping channel with Kraus operators $\widetilde{E}_i^{\textrm{\relsize{-2}{(AD)}}}=e^{-i\Omega(t) \ket{1}\bra{1}} \hat{E}_i^{\textrm{\relsize{-2}{(AD)}}}$. 
This amplitude damping channel is, as the noisy channels considered in the previous sections, generally non-Markovian and includes cases in which the environment consists of a finite set of oscillators (even one).

\section{Conclusions}\label{sec:conclusions}

One-qubit noisy channels (depolarizing, dephasing and amplitude-damping channels) are usually described either by their Kraus representations or by Lindblad master equations with constant decay rates, the typical Markovian master equations. In this paper, simple models for a depolarizing channel have been considered, and models for dephasing and amplitude damping have been briefly reviewed. No classical model for the amplitude-damping channel was found, perhaps because the steady state of this channel is a pure state. 

The most important point raised in this paper is the following. The same Kraus representation corresponds to different dynamical problems, including Markovian master equations with constant decay rates, Markovian master equations with time-dependent non-negative decay rates, and non-Markovian master equations with temporarily negative decay rates. This observation may be useful for analyses of quantum information processes in the most promising scalable realizations of quantum processors, which are often non-Markovian.

\section*{Acknowledgements}
Partial funding for this research was provided by Divisi\'on de Investigaci\'on Sede Bogot\'a - Universidad Nacional de Colombia under project 9366.

\section*{References}
% \bibliographystyle{unsrt}
% \bibliography{/home/karen/Sincronizar/work/mywritings/En_Preparacion/2012/DecoherenceMicroscopicModels/Referencias/Referencias}

\end{document}